\documentclass[conference]{IEEEtran}
\IEEEoverridecommandlockouts
\usepackage{graphicx}
\usepackage{cite}
\usepackage{bm}
\usepackage{amsfonts}
\usepackage{amstext}
\usepackage{amsthm,amsmath,amssymb}
\usepackage{mathrsfs}
\usepackage[mathscr]{eucal}
\usepackage[T1]{fontenc}
\usepackage{booktabs}
\usepackage[noend]{algpseudocode}
\usepackage{multirow}
\usepackage{tabularx}
\usepackage{exscale}
\usepackage{caption}
\usepackage{float} 
\usepackage{graphicx}
\usepackage{subfigure}
\usepackage[nocfg]{nomencl}
\usepackage{framed}
\usepackage{epstopdf}
\usepackage{threeparttable}
\usepackage{booktabs}
\usepackage{tabularx}
\usepackage{xcolor}
\usepackage{threeparttable}
\def\BibTeX{{\rm B\kern-.05em{\sc i\kern-.025em b}\kern-.08em
	T\kern-.1667em\lower.7ex\hbox{E}\kern-.125emX}}

\makeatletter
\newcommand{\linebreakand}{%
\end{@IEEEauthorhalign}
\hfill\mbox{}\par
\mbox{}\hfill\begin{@IEEEauthorhalign}
}
\makeatother

\begin{document}

\title{Replay-guided Test-time Adaptation for Fault Diagnosis Under Unseen Operating Conditions}

\author{
	\IEEEauthorblockN{Yakun Wang}
	\IEEEauthorblockA{\textit{MCC5 Group Shanghai Co. LTD} \\
		Shanghai, China \\
		1460105945@qq.com}
	\and
	\IEEEauthorblockN{Pengyu Han}
	\IEEEauthorblockA{\textit{School of Automation} \\
		\textit{Tsinghua University}\\
		Beijing, China \\
		hpy24@mails.tsinghua.edu.cn}
	\and
	\IEEEauthorblockN{Zeyi Liu}
	\IEEEauthorblockA{\textit{Department of Automation} \\
		\textit{Tsinghua University}\\
		Beijing, China \\
		liuzy21@mails.tsinghua.edu.cn}
	\and
	\IEEEauthorblockN{Xiao He}
	\IEEEauthorblockA{\textit{Department of Automation} \\
		\textit{Tsinghua University}\\
		Beijing, China \\
		hexiao@tsinghua.edu.cn
	}
	\linebreakand
	\IEEEauthorblockN{Dongming Cai}
	\IEEEauthorblockA{\textit{MCC5 Group Shanghai Co. LTD} \\
		Shanghai, China \\
		38874188@qq.com
	}
	\and
	\IEEEauthorblockN{Hongshuo Zhao}
	\IEEEauthorblockA{\textit{MCC5 Group Shanghai Co. LTD} \\
		Shanghai, China \\
		zhs17603218096@163.com
	}

	\thanks{This work was supported in part by National Natural Science Foundation of China under grants 62525308, 624B2087, 62473223, and 52172323, in part by Beijing Natural Science Foundation under grant L241016. (\emph{Corresponding author: Xiao He})
	}
	
}

\maketitle
\begin{abstract}
	In modern industrial systems, machinery frequently operates under dynamic environments with continuously varying loads and speeds. Consequently, deep learning-based fault diagnosis models often suffer from severe performance degradation under unseen operating conditions due to complex data distribution shifts. Since existing methods predominantly rely on static offline training, they lack the capability to dynamically adapt to these continuous variations. To address this issue, an integrated framework combining offline domain generalization (DG) and online test-time adaptation (OTTA) is proposed. Initially, a model with preliminary generalization capability is obtained offline by extracting domain-invariant features via adversarial learning. During the online phase, a dual-memory replay mechanism is developed. 
	By selectively storing high-confidence online pseudo-labeled samples and replaying them with historical offline data, the model facilitates adaptation to changing data distributions and helps reduce forgetting of previously learned knowledge
	Experiments on a real-world motor dataset show that the proposed approach achieves competitive performance under the considered unseen operating conditions.
\end{abstract}



\section{Introduction}
With the rapid development of industrial automation and intelligent manufacturing technologies, the complexity and intelligence level of industrial systems are continuously improving, making the safety and stability of equipment operation increasingly critical \cite{hu2024cadm+}. 
Equipment failures not only cause economic losses but also endanger personnel safety in severe cases; therefore, efficient and reliable fault diagnosis is of great significance. In recent years, deep learning has been widely applied to fault diagnosis due to its excellent feature extraction capabilities, and remarkable progress has been made in improving diagnostic accuracy \cite{MPOS-RVFL,Lu2023b}. However, existing methods are mostly modeled based on idealized assumptions, making it difficult to fully cope with the complexity and uncertainty present in actual industrial environments \cite{Su2022a, Han2023}.

During practical operation, industrial equipment is usually in a state of multiple operating conditions due to the influence of various factors, such as environmental disturbances, load variations, condition switching, and equipment aging. Significant data distribution discrepancies often exist among different operating conditions, leading to a noticeable degradation in model performance in cross-condition scenarios \cite{MCFD}.
Traditional fault diagnosis methods are generally established on the independent and identically distributed (i.i.d.) assumption, where training and testing data are considered to originate from the same distribution. Nevertheless, this assumption rarely holds under multi-condition settings, thereby limiting the generalization capability of the models. Consequently, researching multi-condition fault diagnosis problems is of critical importance.

Recently, attempts have been made to enhance the generalization performance of models under different operating conditions through domain adaptation (DA) \cite{Yang2024, Zhang2024, Choudhary2023} or domain generalization (DG) \cite{Qian2023, Li2023b,Chen2023c} methods, yet certain limitations remain. In practical industrial applications, offline labeled data is inherently limited, and unpredictable operating conditions frequently emerge post-deployment. While existing methods primarily rely on static DG models to combat these complex distribution shifts, their fixed parameters prohibit dynamic adjustments to newly arrived data. Consequently, these static approaches inevitably suffer from performance degradation in continuously changing environments. Therefore, developing a fault diagnosis model capable of continuous adaptive updating under unknown conditions is crucial for enhancing system robustness and practicality \cite{11104131,LI2025130137}.

To overcome the inherent limitations of static offline models, online test-time adaptation (OTTA) has recently emerged as a promising paradigm \cite{wang2025search}. 
Unlike conventional offline strategies, OTTA aims to continuously update the deployed model during the inference phase by leveraging the sequential stream of unlabeled test data. By dynamically adjusting the network parameters in real time, OTTA enables the model to actively track complex data distribution shifts and progressively adapt to newly encountered operating conditions. 
Consequently, integrating an OTTA mechanism into fault diagnosis frameworks offers a potential way to reduce the distribution gap between limited offline data and unpredictable online streams.
Driven by the stringent reliability requirements of modern machinery, this continuous adaptation capability opens up a broad development space for robust fault diagnosis in dynamic industrial environments \cite{10688394}.

To address the distribution shift induced by unknown operating conditions, a novel framework is proposed to achieve continuous model adaptation. In the offline phase, adversarial learning is employed within a DG strategy to extract domain-invariant features, establishing a robust initial model. 
During the online phase, pseudo-label filtering and memory replay mechanisms are introduced to incorporate historical knowledge into the adaptation process using incoming unlabeled data streams.
Therefore, the model can be continuously updated to better accommodate dynamic environmental variations.
The main contributions of this paper are summarized as follows:
\begin{enumerate}
	\item An integrated diagnostic framework combining an offline DG strategy and an OTTA method is proposed to address fault diagnosis under distribution shifts caused by unknown operating conditions.
	Specifically, the offline phase adopts adversarial learning to extract domain-invariant features, establishing a robust initial model. 
	Subsequently, the online phase introduces an update mechanism to leverage newly arrived unlabeled data streams for model adaptation.
	
	\item A dual-memory structure based on a replay mechanism is developed. 
	By constructing memory banks and introducing a replay strategy, information from unlabeled online samples can be incorporated during adaptation while previously learned offline knowledge is retained to some extent.
	Consequently, the proposed strategy helps retain previously learned knowledge and improves the model’s ability to adapt under unseen operating conditions.

	\item Extensive experiments are conducted on a real-world motor dataset. Comparative experiments under unknown operating conditions indicate that the proposed method is effective under the evaluated settings.
	
\end{enumerate}

The remainder of this paper is organized as follows. 
The overall framework and specific implementation procedures of the proposed algorithm are detailed in Section II. 
The experimental settings and results are presented in Section III, where an in-depth analysis and discussion of the findings are also conducted. Finally, the main conclusions of this work are summarized in Section IV.

\section{Proposed Method}
\subsection{Problem Formulation}

Consider a real-time industrial system where the observed input at each time step $t$ is denoted as $x_t \in \mathbb{R}^{1 \times k}$, representing measurements from $k$ sensor channels. Although each sample inherently corresponds to a true fault category, denoted by $y_t$, it is imperative to emphasize that $y_t$ remains strictly unobservable during the online diagnostic process. 

During the offline phase, a fully labeled training dataset is collected from $M$ stable operating conditions, denoted as $\{S_1, S_2, \dots, S_M\}$. Each stable condition $S_m$ contains samples corresponding to $N_f$ fault categories, which can be expressed as $S_m = \{(x_i^m, y_i^m) \mid i = 1, 2, \dots, N_m\}$. 
In the online deployment phase, the system frequently operates under varying conditions induced by load fluctuations, mode switching, and other dynamic factors, thereby generating a continuous stream of strictly unlabeled data. Since only a limited subset of stable conditions can be recorded during the offline data collection process, the data distributions corresponding to newly encountered operating conditions are typically unavailable during training and are thus regarded as unknown conditions. 

Formally, the continuous adaptation problem under unsupervised online scenarios is defined as follows: given the labeled offline data from stable conditions $\{S_1, \dots, S_M\}$, the objective is to learn and dynamically update a diagnostic model $f_{\theta}$ utilizing only the incoming unlabeled samples from unknown operating conditions $S_t$. Furthermore, the model must maintain continuous adaptability as the data distribution evolves over time, such that the expected prediction accuracy concerning the unobserved true labels is maximized:
\begin{equation}
	\max_{\theta} \mathbb{E}_{(x_t, y_t) \sim S_t} \left[ \mathbf{1}\big(f_{\theta}(x_t) = y_t \big) \right],
\end{equation}
where $\mathbf{1}(\cdot)$ is the indicator function.

\subsection{Offline Stage}
In the offline phase, the primary objective is to construct a model capable of extracting condition-invariant features from data collected under multiple operating conditions. To this end, the classical Domain-Adversarial Neural Network (DANN) is adopted as the offline model \cite{DANN}. The model consists of three components: a feature extractor $F$, a fault classifier $G_f$, and a condition classifier $G_c$.
The feature extractor $F$ maps an input sample $x \in \mathbb{R}^{1 \times k}$ to a high-dimensional feature representation $h = F(x)$. The fault classifier $G_f$ receives the feature $h$ and outputs the predicted fault label, while the condition classifier $G_c$ takes the same feature $h$ and predicts the corresponding operating condition. This architecture enables the model to maintain strong fault discriminability while learning features that are invariant to operating conditions.

The training of DANN is based on adversarial learning, aiming to simultaneously minimize the fault classification loss and maximize the condition classification loss, so that the extracted features become insensitive to changes in operating conditions. 
Let $\mathcal{D}^{\text{off}}$ denote the complete offline training dataset, and for each offline sample $x_i \in \mathcal{D}^{\text{off}}$, let $d_i \in {1,2,\dots,M}$ denote its operating-condition label.
Formally, the training objectives are defined by the following cross-entropy losses:
\begin{equation}
	\mathcal{L}_f = -\frac{1}{|\mathcal{D}^{\text{off}}|} \sum_{x_i \in \mathcal{D}^{\text{off}}} \sum_{c=1}^{N_f} \mathbf{1}(y_i = c) \log \hat{y}_i^{(c)},
\end{equation}
and
\begin{equation}
	\mathcal{L}_c = -\frac{1}{|\mathcal{D}^{\text{off}}|} \sum_{x_i \in \mathcal{D}^{\text{off}}} \sum_{m=1}^{M} \mathbf{1}(d_i = m) \log \hat{d}_i^{(m)},
\end{equation}
where $\mathcal{L}_f$ corresponds to the fault classification loss, $\mathcal{L}_c$ corresponds to the condition classification loss, $N_f$ represents the total number of fault categories, and $M$ denotes the total number of operating conditions.

During training, a \textit{Gradient Reversal Layer} (GRL) is applied to propagate the negative gradients from the condition classifier to the feature extractor. In this way, the feature extractor $F$ minimizes the fault classification loss while maximizing the condition classification loss, thereby learning features that are invariant to operating conditions. The overall optimization objective can be expressed as:
\begin{equation}
	\max_{\theta_{G_c}} \min_{\theta_F, \theta_{G_f}} \mathcal{L}_f - \lambda \mathcal{L}_c,
\end{equation}
where $\lambda$ is a trade-off coefficient that balances fault discriminability and condition invariance.
Through this training process, the offline feature extractor $F$ and the fault classifier $G_f$ capture fault-relevant features that are robust to variations in operating conditions, providing a reliable initial model and reference features for the subsequent online adaptive updating stage.

Upon the completion of the DANN model training in the offline phase, memory banks must be initialized for adaptive updating during the online phase. For each fault category under each stable operating condition $S_m$, a subset of samples is randomly selected from $\mathcal{D}^{\text{off}}$ to construct the offline memory bank $\mathcal{M}^{\text{off}}$:
\begin{equation}
	\mathcal{M}^{\text{off}} = \bigcup_{m=1}^{M} \bigcup_{c=1}^{N_f} \left\{ (x_i^{m,c}, y_i^{m,c}) \mid i \in \mathcal{I}_{m,c} \right\},
\end{equation}
where $\mathcal{I}_{m,c}$ denotes the set of sample indices extracted from the $m$-th operating condition and the $c$-th fault category. 
The offline memory bank is utilized to retain the historical knowledge acquired during the offline phase, thereby serving as a reference for subsequent online updates. Subsequently, the online memory bank $\mathcal{M}^{\text{on}}$ is initialized to store high-confidence samples during the online phase to enable sample replay. The initial samples of $\mathcal{M}^{\text{on}}$ are randomly selected from the offline memory bank.

\subsection{Online Stage}

During the online stage, the system continuously receives unlabeled data streams from unknown operating conditions. Assuming an unlabeled online sample $x^t$ arrives at time step $t$, the prediction confidence $\tilde{\mathcal{C}}^{t-1}$ and the pseudo-label $\tilde{y}^{t-1}$ are obtained using the current model parameters updated at the previous time step. To align with the offline architecture, the diagnostic model is denoted as the composition of the feature extractor and the classifier ($\mathcal{H} = G_f \circ F$). The confidence and pseudo-label are calculated as:
\begin{equation}
	\tilde{\mathcal{C}}^{t-1} = \max_{c \in \{1, 2, \dots, N_f\}} \mathcal{H}^{t-1}(x^t)^{(c)}
\end{equation}
\begin{equation}
	\tilde{y}^{t-1} = \mathop{\arg\max}_{c \in \{1, 2, \dots, N_f\}} \mathcal{H}^{t-1}(x^t)^{(c)},
\end{equation}
where $\mathcal{H}^{t-1}(x^t)^{(c)}$ represents the predicted probability for the $c$-th class.

To improve the reliability of online updates, incoming samples are filtered using a confidence-based criterion with a predefined threshold.
An online sample is considered to possess a reliable pseudo-label and is subsequently added to the online memory bank $\mathcal{M}^{\text{on}}$ only when its confidence score $\tilde{\mathcal{C}}^{t-1}$ satisfies the threshold condition. Let $K$ denote the maximum capacity of the online memory bank. When $|\mathcal{M}^{\text{on}}| > K$, a First-In-First-Out (FIFO) strategy is adopted to discard the earliest stored samples. This dynamic updating mechanism ensures that the online memory bank consistently reflects the most recent data distribution while maintaining a constant capacity.

To help balance adaptation to distribution changes and retention of previously learned knowledge during continuous updates, an explicit memory bank replay mechanism is introduced. 
Specifically, a joint memory bank is defined as $\mathcal{M}^{\text{all}} = \mathcal{M}^{\text{off}} \cup \mathcal{M}^{\text{on}}$ to merge the historical offline samples and the recent online samples. Instead of updating the model solely with the incoming data stream, data batches are continuously sampled from the joint memory bank $\mathcal{M}^{\text{all}}$ and replayed into the network for parameter optimization. The cross-entropy loss under this replay mechanism is formulated as:
\begin{equation}
	\mathcal{L}_{\text{ce}} = -\frac{1}{|\mathcal{M}^{\text{all}}|} \sum_{(x_i, \tilde{y}_i) \in \mathcal{M}^{\text{all}}} \sum_{c=1}^{N_f} \mathbf{1}(\tilde{y}_i = c) \log \mathcal{H}(x_i)^{(c)},
\end{equation}
where $\tilde{y}_i$ represents either the ground-truth label for samples originating from $\mathcal{M}^{\text{off}}$ or the assigned pseudo-label for samples drawn from $\mathcal{M}^{\text{on}}$. 

\section{Experiment}
\subsection{Experimental Setup}
\begin{figure}[!h]
\centering
\includegraphics[width=3.4in, keepaspectratio]{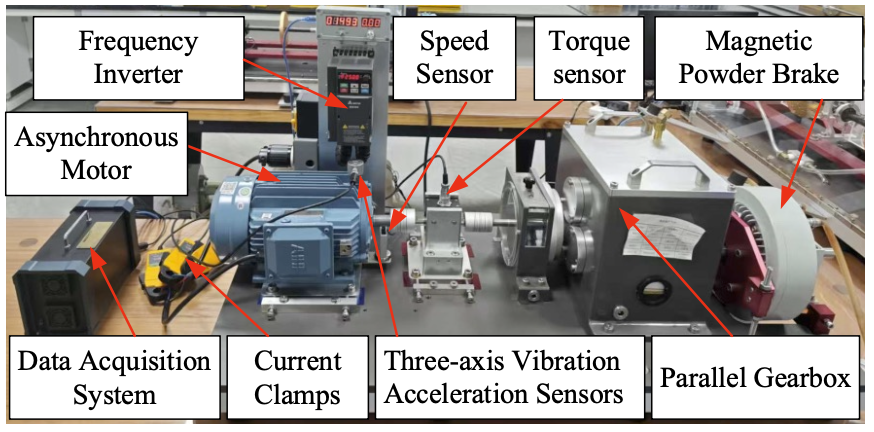}
\caption{The physical configuration of the experimental motor test rig.}
\label{fig_setup}
\end{figure}

In this section, the motor fault diagnosis tasks under diverse operating conditions are investigated. As illustrated in Fig. \ref{fig_setup}, the experimental test rig is composed of a 2.2 kW three-phase asynchronous motor, a torque sensor, a two-stage parallel gearbox, a magnetic powder brake functioning as a load simulator, and a comprehensive measurement and control system. 

\begin{figure*}[!h]
	\centering
	\subfigure[Injection Spot 30\%, Dynamic Eccentricity]{
		\includegraphics[width=0.30\textwidth]{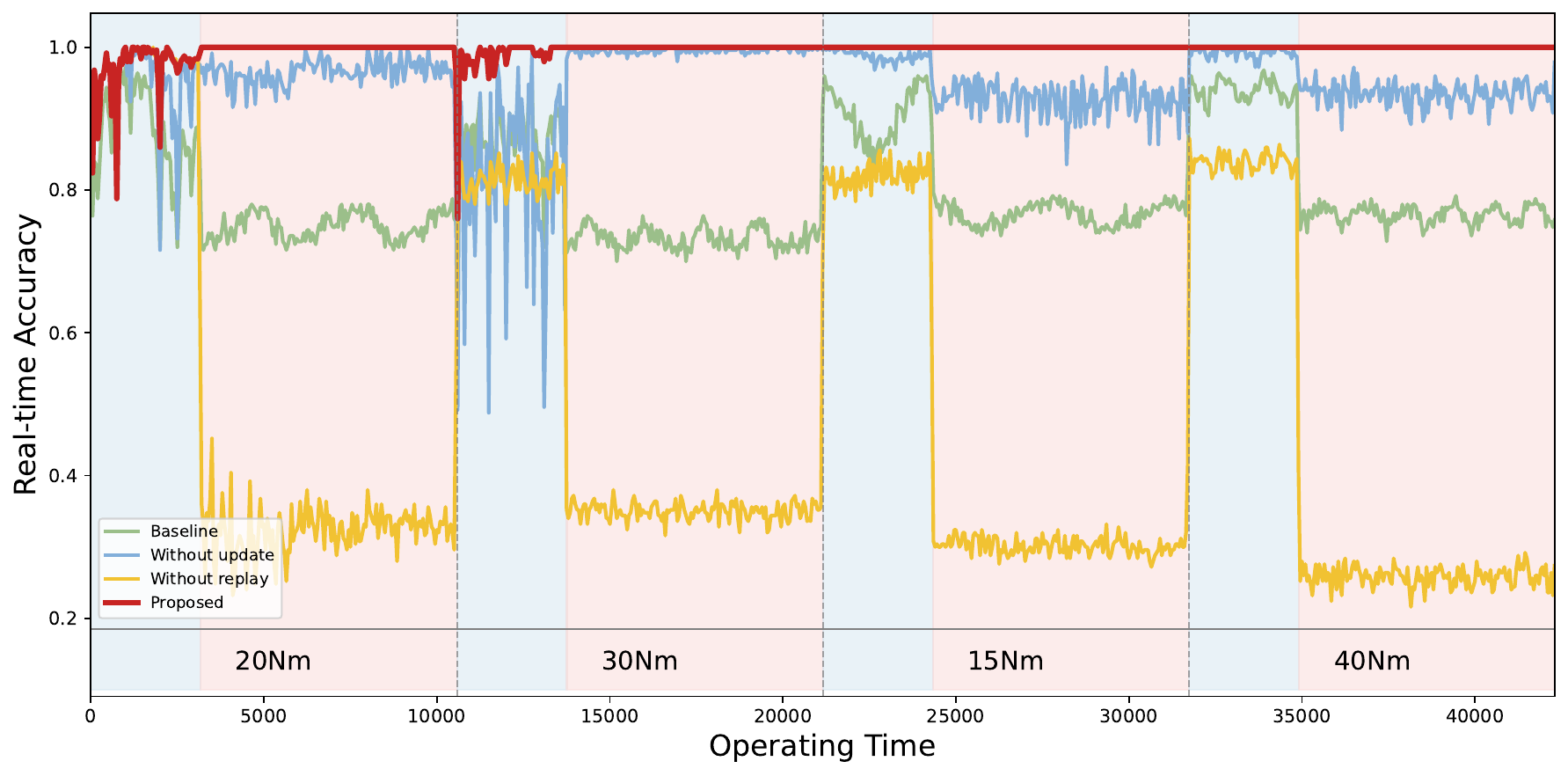}
	}
	\centering
	\subfigure[Injection Spot 50\%, Dynamic Eccentricity]{
		\includegraphics[width=0.30\textwidth]{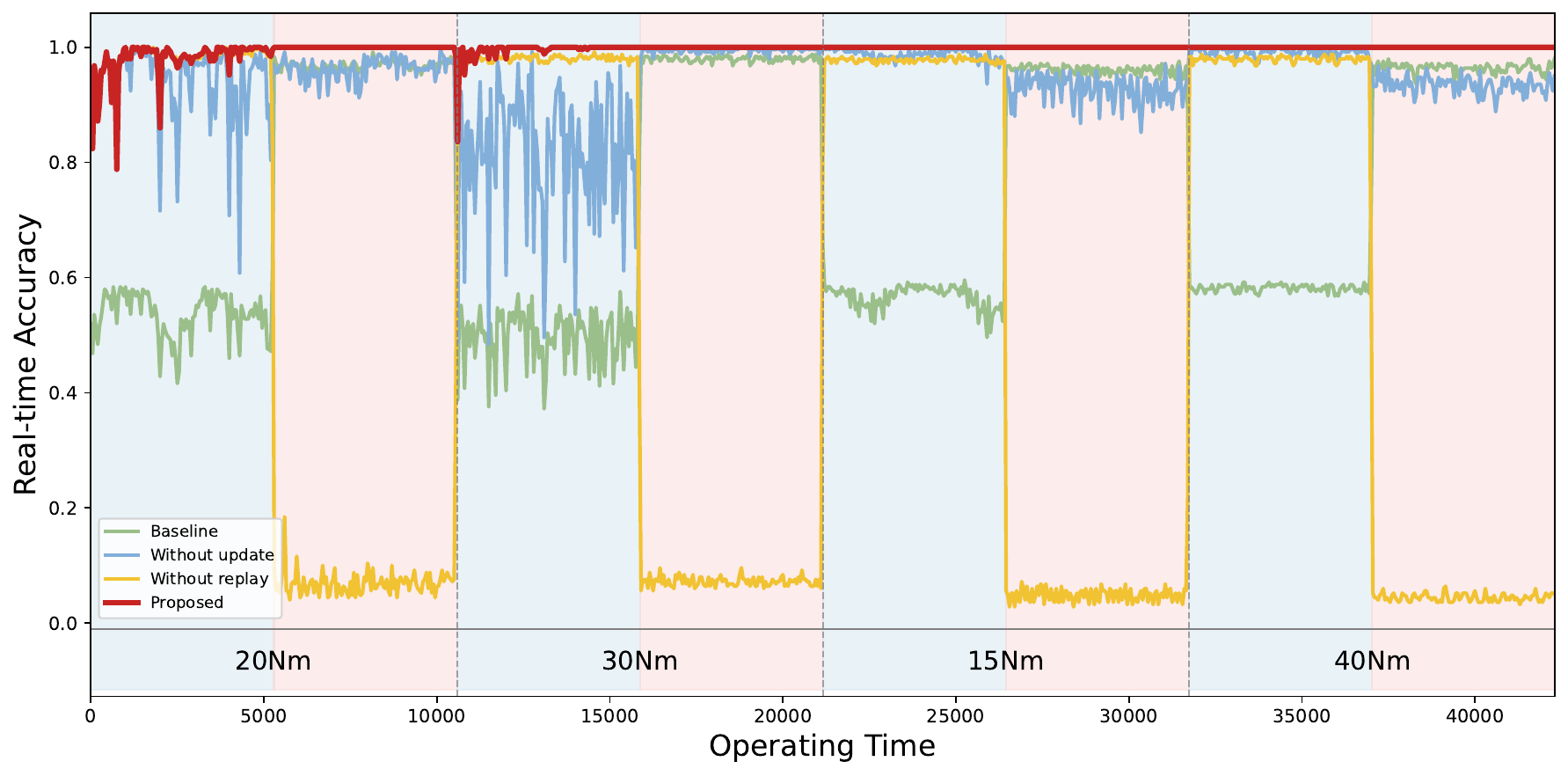}
	}
	\centering
	\subfigure[Injection Spot 70\%, Dynamic Eccentricity]{
		\includegraphics[width=0.30\textwidth]{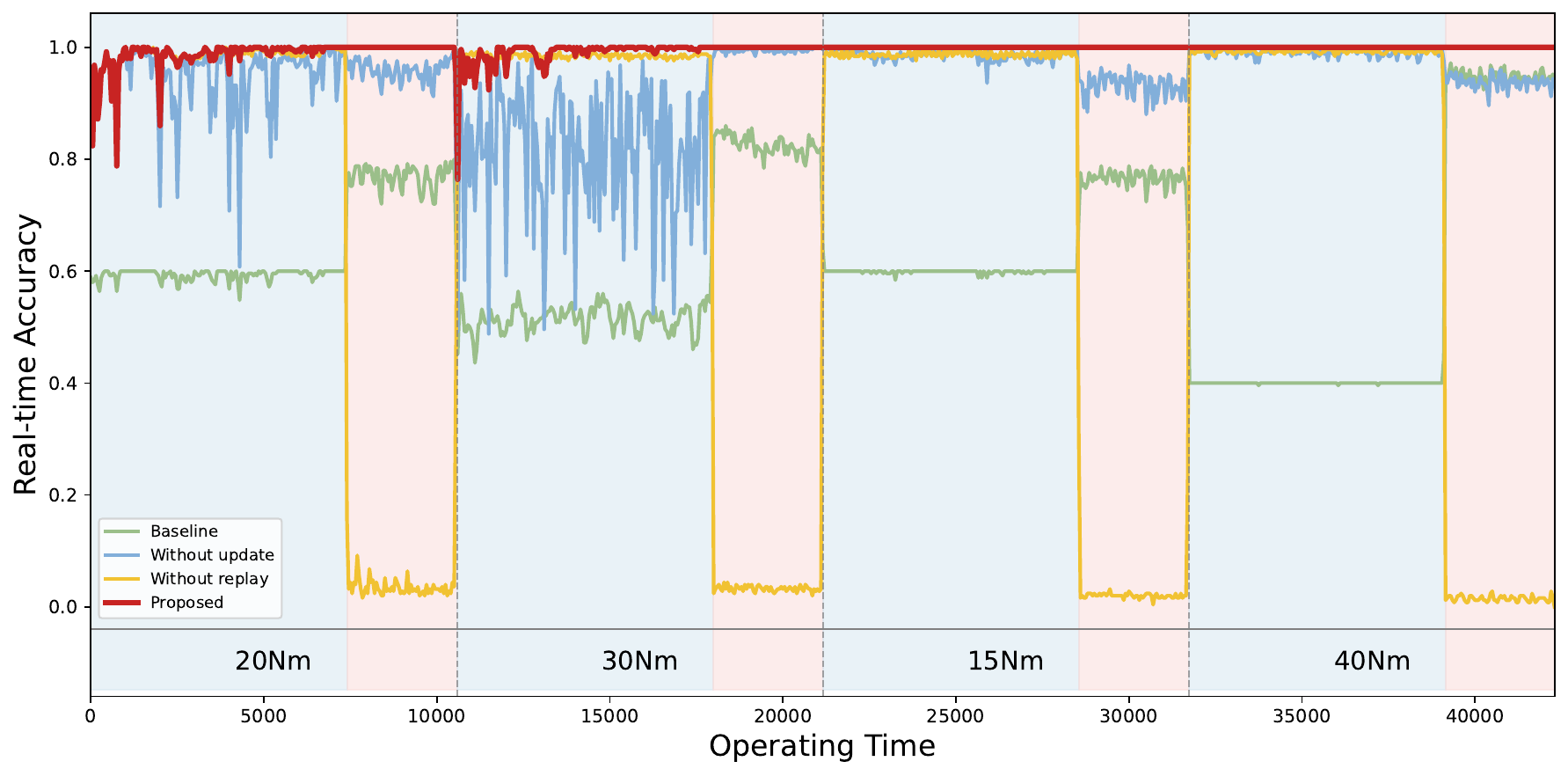}
	}\\
	\centering
	\subfigure[Injection Spot 30\%, Static Eccentricity]{
		\includegraphics[width=0.30\textwidth]{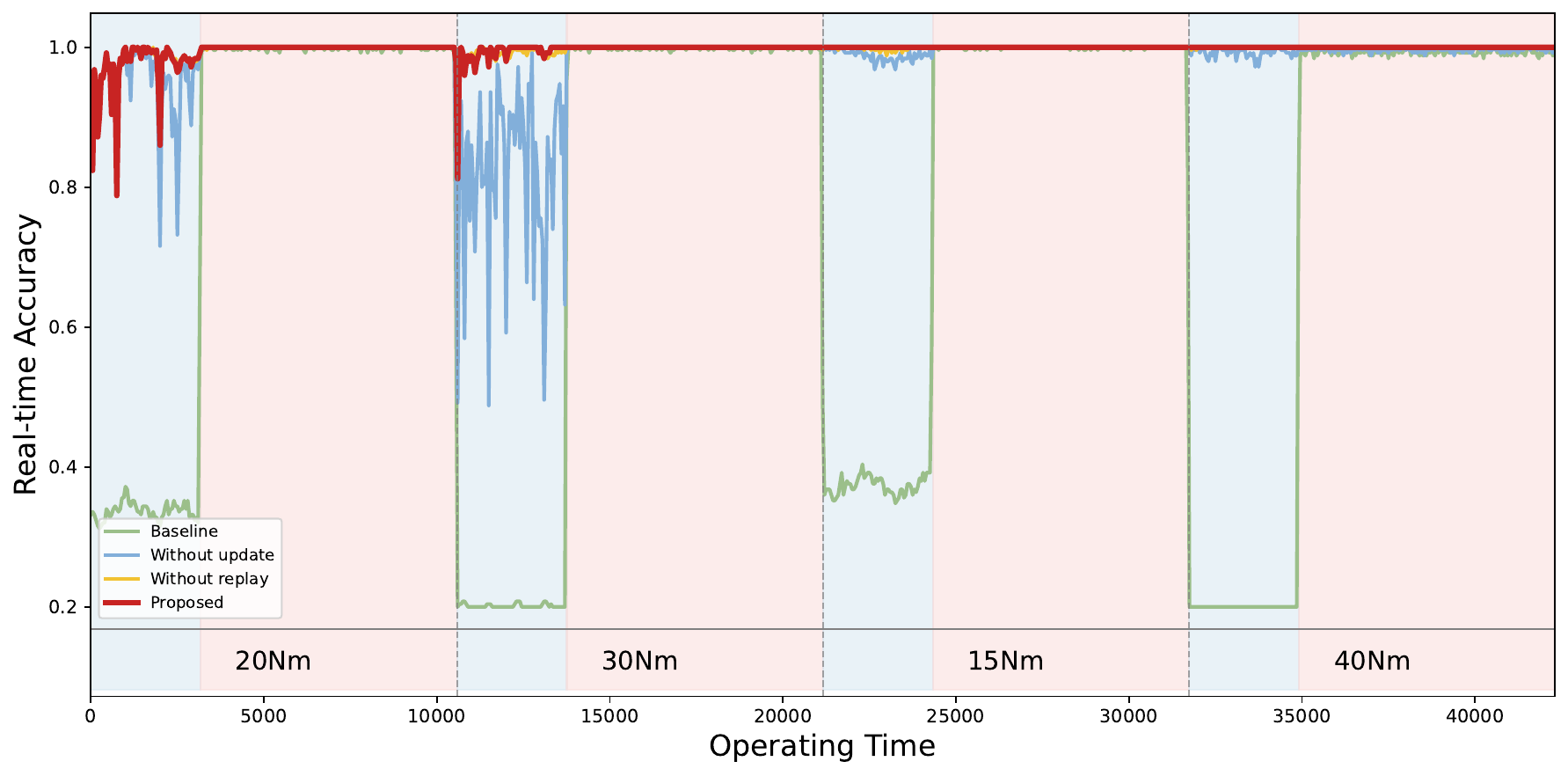}
	}
	\centering
	\subfigure[Injection Spot 50\%, Static Eccentricity]{
		\includegraphics[width=0.30\textwidth]{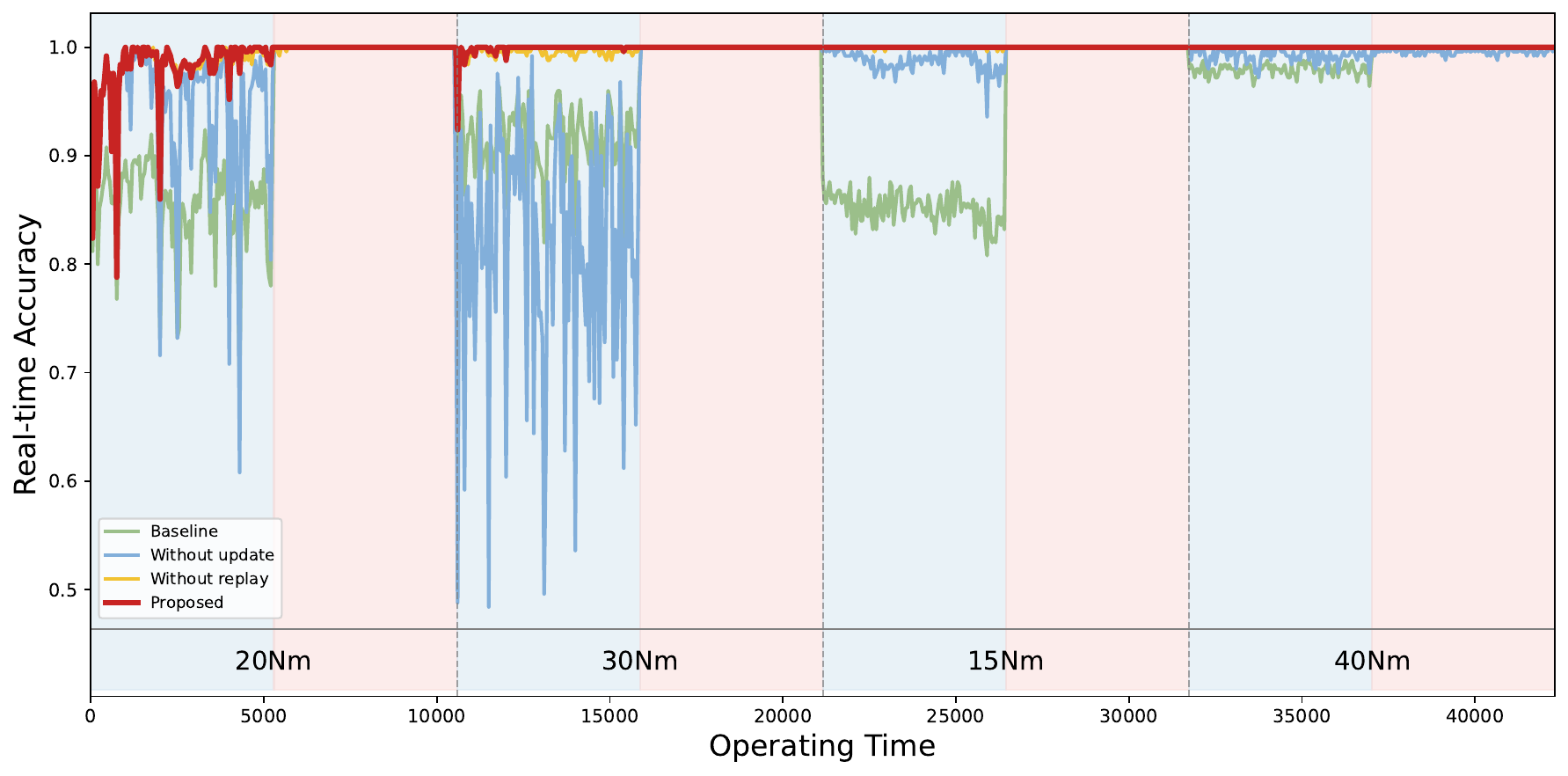}
	}
	\centering
	\subfigure[Injection Spot 70\%, Static Eccentricity]{
		\includegraphics[width=0.30\textwidth]{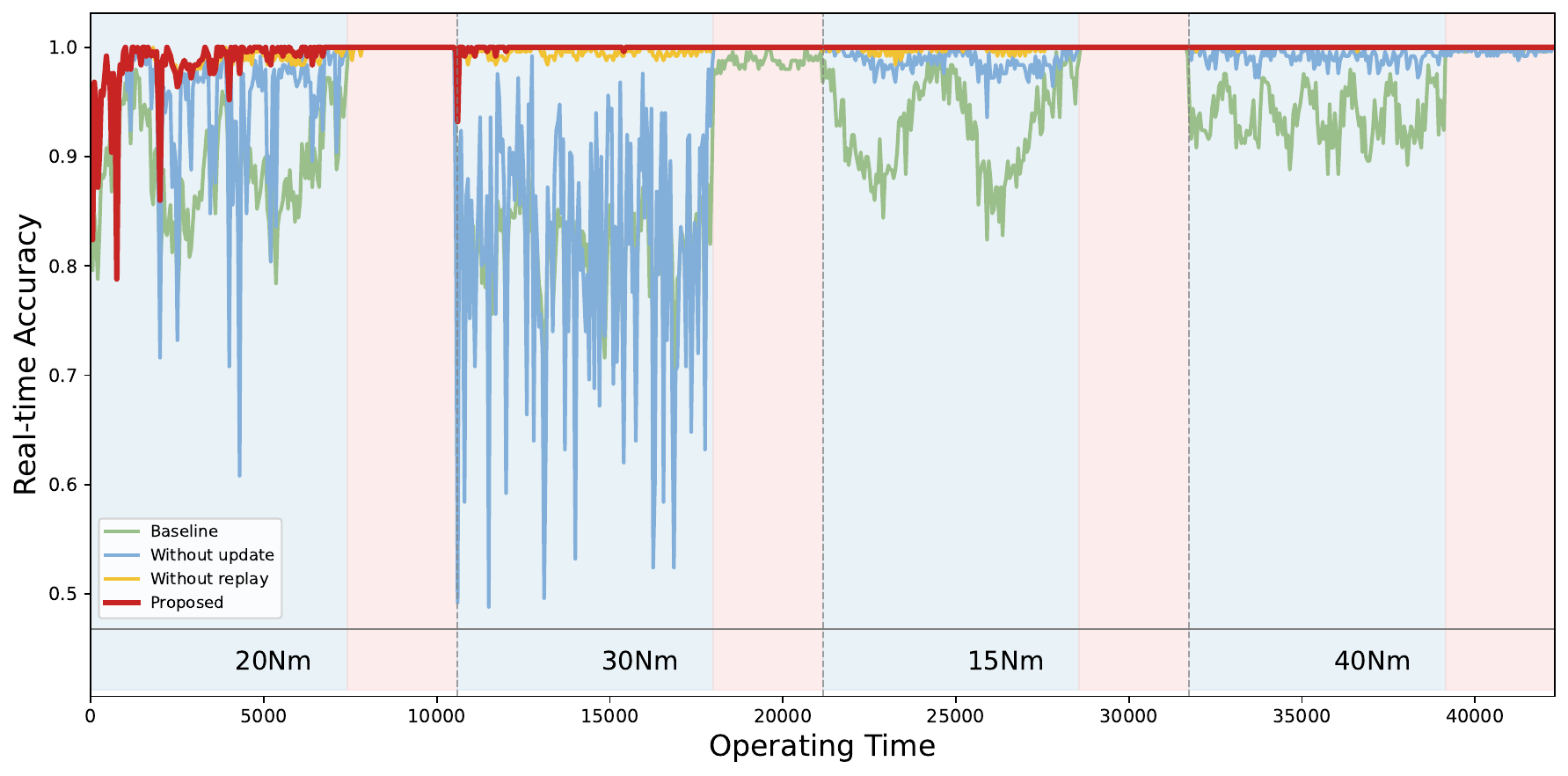}
	}\\
	\centering
	\subfigure[Injection Spot 30\%, Voltage Unbalance]{
		\includegraphics[width=0.30\textwidth]{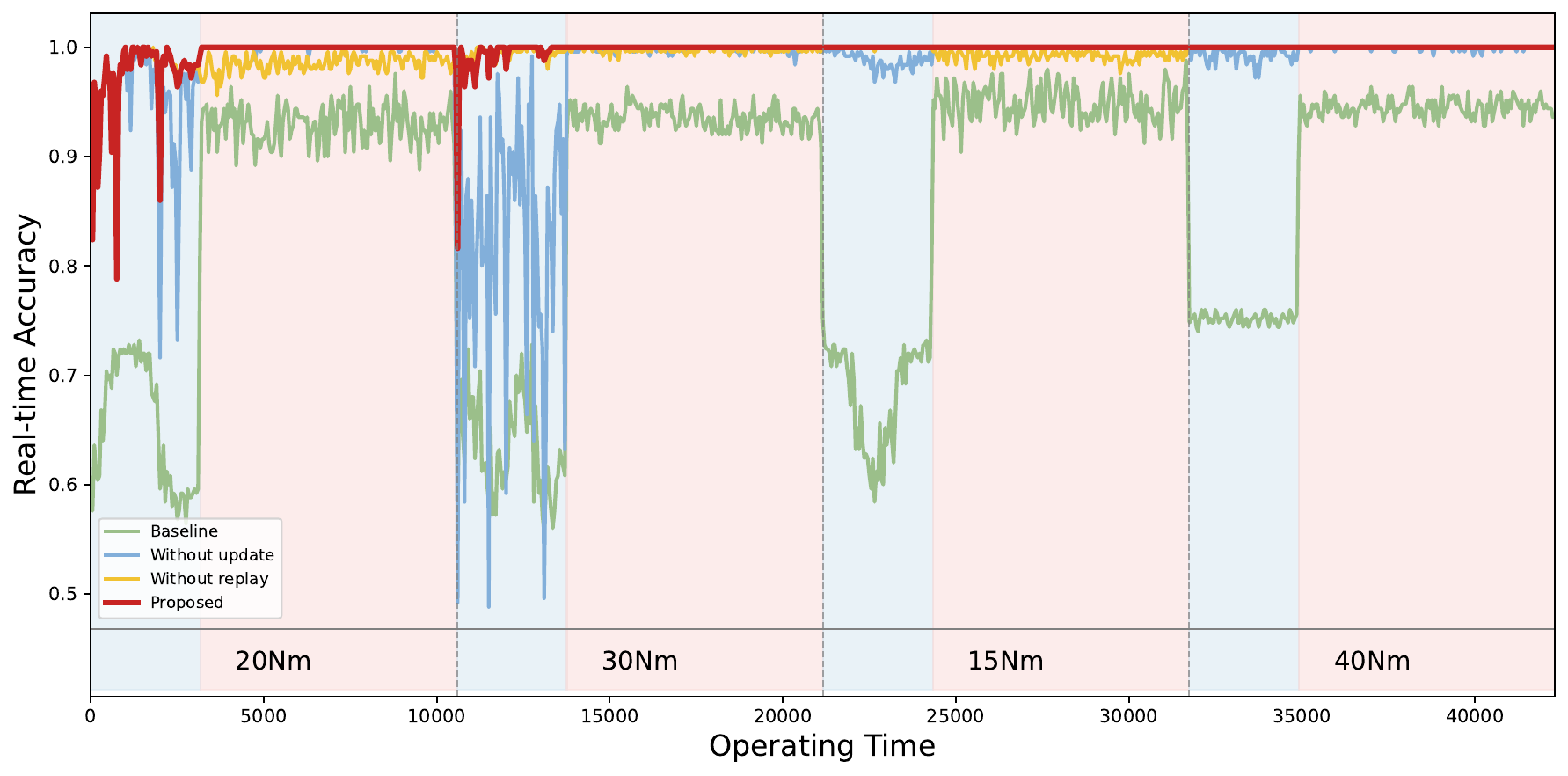}
	}
	\centering
	\subfigure[Injection Spot 50\%, Voltage Unbalance]{
		\includegraphics[width=0.30\textwidth]{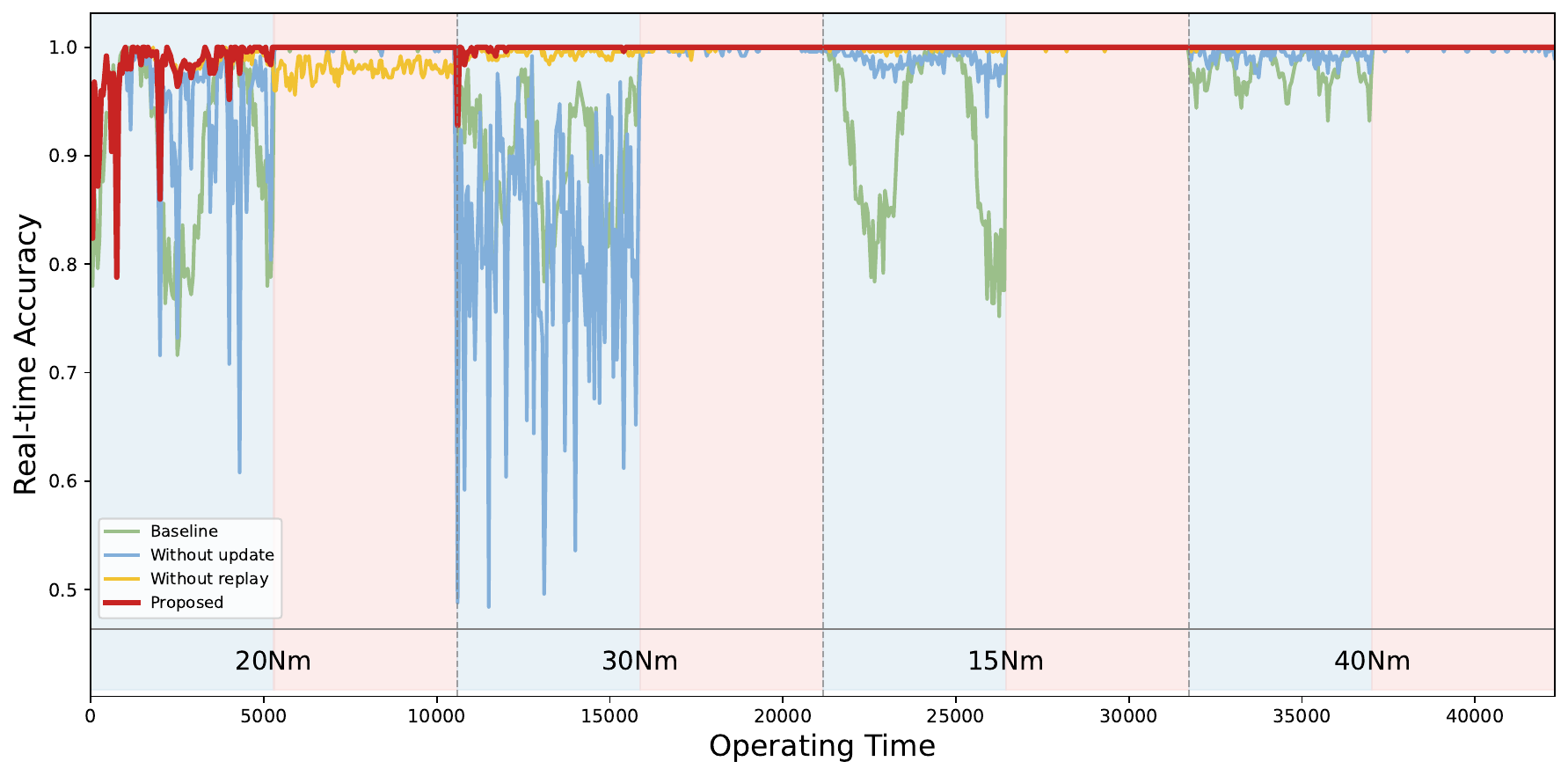}
	}
	\centering
	\subfigure[Injection Spot 70\%, Voltage Unbalance]{
		\includegraphics[width=0.30\textwidth]{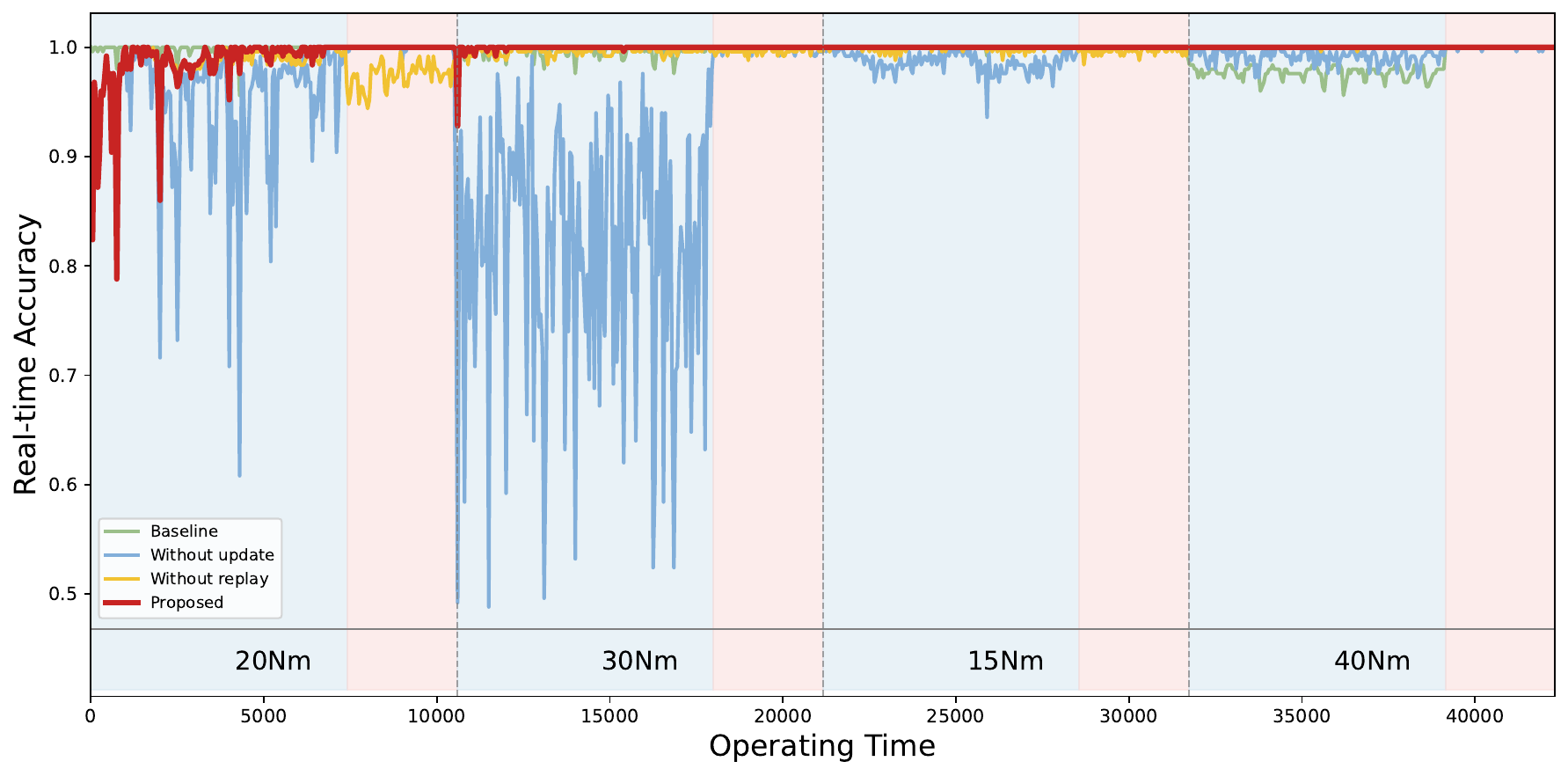}
	}
	
\caption{The learning curves of real-time accuracy for the comparison methods under different fault types and injection spots.}
	\label{CS}
\end{figure*}
An eight-channel synchronous data acquisition system was deployed, encompassing a triaxial vibration acceleration sensor (TES001V, 100 mV/g) at the drive end, three-phase current clamps (Fluke-i30s, 100 mV/A), a torque sensor (S2001, $\pm$0.5\% F.S, 100 mV/Nm), and a key-phase signal. Signals were sampled at 12.8 kHz to satisfy the Nyquist theorem, recording raw voltage data for subsequent metric conversion. Experiments were conducted under 12 steady and transitional operating conditions, with the ambient temperature strictly controlled within $\pm$2$^{\circ}$C. The resulting dataset comprises 282 90-second recordings archived in CSV format. 
A total of 24 fault configurations (including normal states, rotor/stator anomalies, eccentricities, bearing defects, and compound faults) were physically injected via laser etching with a 0.01 mm precision. 
Detailed specifications are provided in \cite{MCC5-motor}.
For subsequent model training, the 6-dimensional raw signals are segmented using a sliding window of length 1024 and a step size of 64. These segments are batched with a size of 128, yielding an input tensor of $\mathbb{R}^{128 \times 6144}$ per iteration.

\begin{table*}[!h]
	\centering
\caption{Diagnostic performance of all methods across various experimental settings.}
	\label{tab:ablation}
	\setlength{\tabcolsep}{3mm}{
		\begin{tabular}{c|l|cccccc}
			\specialrule{0.1em}{3pt}{1pt}
			\specialrule{0.1em}{1pt}{1pt}
			\multirow{2.5}{*}{Injection Spot} & \multirow{2.5}{*}{Fault Type} & \multicolumn{2}{c}{\multirow{2.5}{*}{Operation Condition}} & \multicolumn{4}{c}{Method} \\
			\cmidrule{5-8}
			& & \multicolumn{2}{c}{} & Baseline & Without Update & Without Replay & \textbf{Proposed} \\
			\midrule
			\multirow{12}[12]{*}{30\%} & \multirow{4}[4]{*}{Dynamic Eccentricity} & \multirow{2}[2]{*}{UC} & 20Nm & 0.788$\pm$0.266 & 0.960$\pm$0.054 & 0.519$\pm$0.206 & \textbf{0.991$\pm$0.010} \\
			&  &  & 30Nm & 0.772$\pm$0.264 & 0.943$\pm$0.032 & 0.489$\pm$0.173 & \textbf{0.996$\pm$0.002} \\
			\cmidrule{3-8}
			&  & \multirow{2}[2]{*}{KC} & 15Nm & 0.808$\pm$0.271 & 0.945$\pm$0.072 & 0.457$\pm$0.146 & \textbf{1.000$\pm$0.000} \\
			&  &  & 40Nm & 0.818$\pm$0.270 & 0.952$\pm$0.087 & 0.433$\pm$0.125 & \textbf{1.000$\pm$0.000} \\
			\cmidrule{2-8}
			& \multirow{4}[4]{*}{Static Eccentricity} & \multirow{2}[2]{*}{UC} & 20Nm & 0.800$\pm$0.115 & 0.984$\pm$0.014 & \textbf{0.992$\pm$0.010} & 0.991$\pm$0.010 \\
			&  &  & 30Nm & 0.760$\pm$0.118 & 0.946$\pm$0.029 & \textbf{0.998$\pm$0.005} & 0.997$\pm$0.002 \\
			\cmidrule{3-8}
			&  & \multirow{2}[2]{*}{KC} & 15Nm & 0.812$\pm$0.117 & 0.997$\pm$0.007 & 1.000$\pm$0.001 & \textbf{1.000$\pm$0.000} \\
			&  &  & 40Nm & 0.756$\pm$0.111 & 0.995$\pm$0.006 & 1.000$\pm$0.000 & \textbf{1.000$\pm$0.000} \\
			\cmidrule{2-8}
			& \multirow{4}[4]{*}{Voltage Unbalance} & \multirow{2}[2]{*}{UC} & 20Nm & 0.847$\pm$0.096 & 0.984$\pm$0.013 & 0.982$\pm$0.019 & \textbf{0.991$\pm$0.010} \\
			&  &  & 30Nm & 0.845$\pm$0.101 & 0.945$\pm$0.028 & 0.997$\pm$0.004 & \textbf{0.997$\pm$0.002} \\
			\cmidrule{3-8}
			&  & \multirow{2}[2]{*}{KC} & 15Nm & 0.869$\pm$0.107 & 0.997$\pm$0.007 & 0.995$\pm$0.011 & \textbf{1.000$\pm$0.000} \\
			&  &  & 40Nm & 0.887$\pm$0.121 & 0.997$\pm$0.004 & \textbf{1.000$\pm$0.000} & \textbf{1.000$\pm$0.000} \\
			\midrule
			\multirow{12}[12]{*}{50\%} & \multirow{4}[4]{*}{Dynamic Eccentricity} & \multirow{2}[2]{*}{UC} & 20Nm & 0.752$\pm$0.211 & 0.952$\pm$0.050 & 0.524$\pm$0.025 & \textbf{0.990$\pm$0.012} \\
			&  &  & 30Nm & 0.739$\pm$0.197 & 0.905$\pm$0.053 & 0.526$\pm$0.030 & \textbf{0.998$\pm$0.002} \\
			\cmidrule{3-8}
			&  & \multirow{2}[2]{*}{KC} & 15Nm & 0.764$\pm$0.221 & 0.958$\pm$0.057 & 0.513$\pm$0.016 & \textbf{1.000$\pm$0.000} \\
			&  &  & 40Nm & 0.773$\pm$0.228 & 0.963$\pm$0.066 & 0.512$\pm$0.015 & \textbf{1.000$\pm$0.000} \\
			\cmidrule{2-8}
			& \multirow{4}[4]{*}{Static Eccentricity} & \multirow{2}[2]{*}{UC} & 20Nm & 0.927$\pm$0.088 & 0.970$\pm$0.022 & 0.990$\pm$0.013 & \textbf{0.990$\pm$0.012} \\
			&  &  & 30Nm & 0.957$\pm$0.028 & 0.908$\pm$0.051 & 0.998$\pm$0.005 & \textbf{0.999$\pm$0.001} \\
			\cmidrule{3-8}
			&  & \multirow{2}[2]{*}{KC} & 15Nm & 0.926$\pm$0.107 & 0.994$\pm$0.012 & 1.000$\pm$0.000 & \textbf{1.000$\pm$0.000} \\
			&  &  & 40Nm & 0.990$\pm$0.019 & 0.994$\pm$0.008 & \textbf{1.000$\pm$0.000} & \textbf{1.000$\pm$0.000} \\
			\cmidrule{2-8}
			& \multirow{4}[4]{*}{Voltage Unbalance} & \multirow{2}[2]{*}{UC} & 20Nm & 0.954$\pm$0.052 & 0.970$\pm$0.022 & 0.980$\pm$0.020 & \textbf{0.990$\pm$0.012} \\
			&  &  & 30Nm & 0.948$\pm$0.060 & 0.907$\pm$0.050 & 0.997$\pm$0.005 & \textbf{0.999$\pm$0.001} \\
			\cmidrule{3-8}
			&  & \multirow{2}[2]{*}{KC} & 15Nm & 0.959$\pm$0.048 & 0.994$\pm$0.012 & 0.999$\pm$0.001 & \textbf{1.000$\pm$0.000} \\
			&  &  & 40Nm & 0.990$\pm$0.015 & 0.996$\pm$0.008 & 1.000$\pm$0.000 & \textbf{1.000$\pm$0.000} \\
			\midrule
			\multirow{12}[12]{*}{70\%} & \multirow{4}[4]{*}{Dynamic Eccentricity} & \multirow{2}[2]{*}{UC} & 20Nm & 0.646$\pm$0.300 & 0.952$\pm$0.046 & 0.699$\pm$0.000 & \textbf{0.989$\pm$0.013} \\
			&  &  & 30Nm & 0.608$\pm$0.299 & 0.870$\pm$0.071 & 0.699$\pm$0.005 & \textbf{0.993$\pm$0.004} \\
			\cmidrule{3-8}
			&  & \multirow{2}[2]{*}{KC} & 15Nm & 0.649$\pm$0.302 & 0.970$\pm$0.043 & 0.697$\pm$0.006 & \textbf{1.000$\pm$0.000} \\
			&  &  & 40Nm & 0.564$\pm$0.325 & 0.975$\pm$0.046 & 0.699$\pm$0.002 & \textbf{1.000$\pm$0.000} \\
			\cmidrule{2-8}
			& \multirow{4}[4]{*}{Static Eccentricity} & \multirow{2}[2]{*}{UC} & 20Nm & 0.921$\pm$0.103 & 0.964$\pm$0.028 & 0.988$\pm$0.016 & \textbf{0.989$\pm$0.013} \\
			&  &  & 30Nm & 0.864$\pm$0.134 & 0.871$\pm$0.071 & 0.996$\pm$0.007 & \textbf{0.999$\pm$0.001} \\
			\cmidrule{3-8}
			&  & \multirow{2}[2]{*}{KC} & 15Nm & 0.949$\pm$0.079 & 0.991$\pm$0.017 & 0.999$\pm$0.003 & \textbf{1.000$\pm$0.000} \\
			&  &  & 40Nm & 0.955$\pm$0.057 & 0.993$\pm$0.011 & 1.000$\pm$0.000 & \textbf{1.000$\pm$0.000} \\
			\cmidrule{2-8}
			& \multirow{4}[4]{*}{Voltage Unbalance} & \multirow{2}[2]{*}{UC} & 20Nm & \textbf{0.999$\pm$0.002} & 0.964$\pm$0.028 & 0.981$\pm$0.019 & 0.989$\pm$0.013 \\
			&  &  & 30Nm & 0.998$\pm$0.004 & 0.871$\pm$0.070 & 0.996$\pm$0.005 & \textbf{0.999$\pm$0.001} \\
			\cmidrule{3-8}
			&  & \multirow{2}[2]{*}{KC} & 15Nm & 1.000$\pm$0.000 & 0.991$\pm$0.017 & 0.998$\pm$0.002 & \textbf{1.000$\pm$0.000} \\
			&  &  & 40Nm & 0.984$\pm$0.032 & 0.994$\pm$0.012 & 1.000$\pm$0.000 & \textbf{1.000$\pm$0.000} \\
			\specialrule{0.1em}{3pt}{1pt}
			\specialrule{0.1em}{1pt}{1pt}
		\end{tabular}
		\begin{tablenotes}
			\footnotesize
			\item[]Notes: UC denotes unknown operating conditions. KC denotes known operating conditions.
	\end{tablenotes}}
	\label{tab:addlabel}%
\end{table*}%

\subsection{Experimental Setup}

The subsequent experimental evaluation encompasses an offline training phase using 15 Nm and 40 Nm loads, allocating 556 samples per fault category for each condition. The continuous online adaptation phase is then evaluated sequentially across 20 Nm, 30 Nm, 15 Nm, and 40 Nm conditions, comprising a continuous data stream of 10,580 samples per condition. Specifically, the unseen conditions (20 Nm, 30 Nm) are utilized to evaluate the model's adaptability to novel environments, whereas the seen offline conditions (15 Nm, 40 Nm) assess its resistance to catastrophic forgetting.
Within each stable operating condition, a specific fault is injected at a predetermined moment. Specifically, samples preceding the injection point represent healthy states, whereas those succeeding it constitute continuous fault states. Three distinct fault types are considered in the experiments. To comprehensively evaluate the robustness of the model, three different fault injection positions are configured at 30\%, 50\%, and 70\% of the continuous data sequence for each condition.

A Multi-Layer Perceptron (MLP) architecture is adopted as the backbone for the diagnostic model. The feature extractor $F$ maps the raw input signal into a 128-dimensional feature representation using four hidden layers, with a neuron configuration of 6144-1024-512-256-128. Based on these extracted features, the fault classifier $G_f$ is composed of a two-layer MLP with 128-32-$N_f$ neurons, where $N_f$ denotes the number of fault categories, outputting the diagnostic logits. The domain discriminator $G_c$ is constructed using a four-layer MLP with 128-128-128-64-$M$ neurons, where $M$ represents the number of source domains in the offline phase. 

For the adversarial training in DANN, the adaptation penalty parameter $\lambda$ is scheduled to gradually increase from 0 to 1 as the training progresses. The Adam optimizer is employed for network optimization. The learning rate is set to $10^{-3}$ for the offline phase and reduced to $10^{-4}$ for the online adaptation phase. The offline model is trained for 50 epochs, while each online update is performed for 30 epochs. Furthermore, to facilitate the memory replay mechanism, the offline memory bank is constructed by allocating 100 samples per fault category for each condition, and the capacity of the online memory bank queue is fixed at 1024. To ensure statistical reliability, the experimental results of all methods are averaged over five independent trials.

\subsection{Comparison Study}
To evaluate the proposed approach, comparative experiments are conducted against several baseline methods. These include: 1) \textbf{Baseline}, which is a standard MLP model trained solely with cross-entropy loss without any domain adaptation strategies; 2) \textbf{Without Update}, representing the DANN model trained only in the offline phase and directly applied to the online data stream without test-time adaptation; 3) \textbf{Without Replay}, where online adaptation is performed utilizing only the newly acquired samples in the online memory bank ($\mathcal{M}^{\text{on}}$), without incorporating offline historical samples during the update process; and 4) \textbf{Proposed Method}, referring to the complete continuous adaptation framework equipped with the dual-memory replay mechanism proposed in this paper.

The diagnostic results, represented as the average accuracy and standard deviation (Acc $\pm$ SD) over five independent trials, are summarized in Table \ref{tab:ablation}. 
An analysis of the experimental data reveals that the Baseline model exhibits the poorest overall performance and the highest standard deviations, particularly under dynamic eccentricity faults. 
This indicates that a model trained purely on source-domain data, without additional generalization or adaptation strategies, may have limited ability to cope with the severe distribution shifts caused by varying loads and fault injection timings.

Moreover, the necessity of the online update mechanism is explicitly verified by the performance of the Without Update method. Although it demonstrates decent performance under known conditions (KC) due to the offline adversarial training, noticeable performance degradation is experienced under unseen conditions (UC). This degradation is especially pronounced when the fault injection point is delayed (e.g., at the 70\% injection spot under 30 Nm, the accuracy drops to 0.871$\pm$0.071). 
These results suggest that static models may struggle to track continuous distribution changes in online scenarios.

The necessity of the replay mechanism is evidenced by the severe performance collapse of the Without Replay method, where accuracies drop drastically to 0.433-0.526 under dynamic eccentricity faults. 
These results suggest that relying only on recent online samples may lead to the loss of previously learned knowledge.
Conversely, the proposed replay formulation utilizes the offline memory bank as a structural anchor. Rehearsing these anchor samples preserves the robust discriminative capabilities acquired offline, explicitly preventing catastrophic forgetting. Simultaneously, replaying dynamically updated online samples drives the model to progressively adapt to ongoing distribution shifts, thereby achieving robust continuous diagnosis under unknown conditions.

In contrast, the Proposed Method achieves competitive or superior diagnostic performance in most evaluated settings, often yielding high accuracies and relatively stable results across different fault types, injection spots, and operating conditions.
By integrating offline domain generalization with an online dual-memory replay strategy, the proposed framework shows the potential to adapt to unseen operating conditions while helping retain previously learned diagnostic knowledge during continuous fault diagnosis.

\section{Conclusion}
In this paper, an integrated framework combining offline domain generalization and online test-time adaptation has been proposed to address the critical challenge of fault diagnosis under unseen operating conditions. To enable the model to dynamically adapt to continuous data distribution shifts, a dual-memory structure incorporating a sample replay strategy has been developed. 
By selectively storing high-confidence online pseudo-labeled samples and replaying them alongside offline historical data, the proposed method helps retain previously learned knowledge during online adaptation.
Consequently, the adaptation capability of the diagnostic model is improved under the evaluated settings.
Furthermore, experimental results on the collected motor dataset suggest that the proposed approach is a promising option for online fault diagnosis under unseen operating conditions.

\bibliographystyle{ieeetr}
\bibliography{MCFD_250406}

@article{Chen2023c,
  title = {A {{Meta-Learning Method}} for {{Electric Machine Bearing Fault Diagnosis Under Varying Working Conditions With Limited Data}}},
  author = {Chen, Jianjun and Hu, Weihao and Cao, Di and Zhang, Zhenyuan and Chen, Zhe and Blaabjerg, Frede},
  year = {2023},
  month = mar,
  journal = {IEEE Transactions on Industrial Informatics},
  volume = {19},
  number = {3},
  pages = {2552--2564},
  issn = {1551-3203, 1941-0050},
  doi = {10.1109/TII.2022.3165027},
  urldate = {2024-01-21},
  lccn = {1}
}

@article{Choudhary2023,
  title = {Passive {{Thermography Based Bearing Fault Diagnosis Using Transfer Learning With Varying Working Conditions}}},
  author = {Choudhary, Anurag and Mian, Tauheed and Fatima, Shahab and Panigrahi, B. K.},
  year = {2023},
  month = mar,
  journal = {IEEE Sensors Journal},
  volume = {23},
  number = {5},
  pages = {4628--4637},
  issn = {1530-437X, 1558-1748, 2379-9153},
  doi = {10.1109/JSEN.2022.3164430},
  urldate = {2024-09-25},
  copyright = {https://ieeexplore.ieee.org/Xplorehelp/downloads/license-information/IEEE.html},
  lccn = {2}
}

@article{Han2023,
  title = {Intelligent Fault Diagnosis of Planetary Gearboxes under Time-Varying Conditions Based on Dynamic Adversarial Balance Adaptation with Multi-Label Information Confusion},
  author = {Han, Songjun and Feng, Zhipeng},
  year = {2023},
  month = jun,
  journal = {Measurement Science and Technology},
  volume = {34},
  number = {6},
  pages = {065014},
  issn = {0957-0233, 1361-6501},
  doi = {10.1088/1361-6501/acc34a},
  urldate = {2024-10-11},
  langid = {american},
  lccn = {3}
}

@article{Li2023b,
  title = {Cross-Domain Augmentation Diagnosis: {{An}} Adversarial Domain-Augmented Generalization Method for Fault Diagnosis under Unseen Working Conditions},
  shorttitle = {Cross-Domain Augmentation Diagnosis},
  author = {Li, Qi and Chen, Liang and Kong, Lin and Wang, Dong and Xia, Min and Shen, Changqing},
  year = {2023},
  month = jun,
  journal = {Reliability Engineering \& System Safety},
  volume = {234},
  pages = {109171},
  issn = {09518320},
  doi = {10.1016/j.ress.2023.109171},
  urldate = {2024-09-25},
  langid = {english},
  lccn = {1}
}

@article{Lu2023b,
  title = {Category-Aware Dual Adversarial Domain Adaptation Model for Rolling Bearings Fault Diagnosis under Variable Conditions},
  author = {Lu, Xingchi and Xu, Weiyang and Jiang, Quansheng and Shen, Yehu and Xu, Fengyu and Zhu, Qixin},
  year = {2023},
  month = sep,
  journal = {Measurement Science and Technology},
  volume = {34},
  number = {9},
  pages = {095104},
  issn = {0957-0233, 1361-6501},
  doi = {10.1088/1361-6501/acd6ac},
  urldate = {2024-10-11},
  langid = {american},
  lccn = {3}
}

@article{Qian2023,
  title = {Relationship {{Transfer Domain Generalization Network}} for {{Rotating Machinery Fault Diagnosis Under Different Working Conditions}}},
  author = {Qian, Quan and Zhou, Jianghong and Qin, Yi},
  year = {2023},
  month = sep,
  journal = {IEEE Transactions on Industrial Informatics},
  volume = {19},
  number = {9},
  pages = {9898--9908},
  issn = {1551-3203, 1941-0050},
  doi = {10.1109/TII.2022.3232842},
  urldate = {2024-09-25},
  copyright = {https://ieeexplore.ieee.org/Xplorehelp/downloads/license-information/IEEE.html},
  langid = {american},
  lccn = {1}
}

@article{Su2022a,
  title = {A Novel Method Based on Deep Transfer Unsupervised Learning Network for Bearing Fault Diagnosis under Variable Working Condition of Unequal Quantity},
  author = {Su, Hao and Yang, Xin and Xiang, Ling and Hu, Aijun and Xu, Yonggang},
  year = {2022},
  month = apr,
  journal = {Knowledge-Based Systems},
  volume = {242},
  pages = {108381},
  issn = {09507051},
  doi = {10.1016/j.knosys.2022.108381},
  urldate = {2024-09-25},
  langid = {english},
  lccn = {1}
}

@article{Yang2024,
  title = {{{PSNN-TADA}}: {{Prototype}} and {{Stochastic Neural Network-Based Twice Adversarial Domain Adaptation}} for {{Fault Diagnosis Under Varying Working Conditions}}},
  shorttitle = {{{PSNN-TADA}}},
  author = {Yang, Xilin and Yuan, Xianfeng and Ye, Tianyi and Zhu, Weijie and Zhou, Fengyu and Jin, Jiong},
  year = {2024},
  journal = {IEEE Transactions on Instrumentation and Measurement},
  volume = {73},
  pages = {1--12},
  issn = {0018-9456, 1557-9662},
  doi = {10.1109/TIM.2023.3338678},
  urldate = {2024-10-11},
  copyright = {https://ieeexplore.ieee.org/Xplorehelp/downloads/license-information/IEEE.html},
  langid = {american},
  lccn = {2}
}

@article{Zhang2024,
  title = {Multi-Modal Data Cross-Domain Fusion Network for Gearbox Fault Diagnosis under Variable Operating Conditions},
  author = {Zhang, Yongchao and Ding, Jinliang and Li, Yongbo and Ren, Zhaohui and Feng, Ke},
  year = {2024},
  month = jul,
  journal = {Engineering Applications of Artificial Intelligence},
  volume = {133},
  pages = {108236},
  issn = {09521976},
  doi = {10.1016/j.engappai.2024.108236},
  urldate = {2024-10-11},
  langid = {english},
  lccn = {2}
}

@ARTICLE{MCFD,
	author={Han, Pengyu and Liu, Zeyi and He, Xiao and Ding, Steven X. and Zhou, Donghua},
	journal={IEEE Transactions on Automation Science and Engineering}, 
	title={Multi-Condition Fault Diagnosis of Dynamic Systems: A Survey, Insights, and Prospects}, 
	year={2025},
	volume={22},
	number={},
	pages={15728-15744},
	doi={10.1109/TASE.2025.3571516}}

@article{LI2025130137,
	title = {A dynamic anchor-based online semi-supervised learning approach for fault diagnosis under variable operating conditions},
	journal = {Neurocomputing},
	volume = {638},
	pages = {130137},
	year = {2025},
	issn = {0925-2312},
	doi = {https://doi.org/10.1016/j.neucom.2025.130137},
	author = {Wei Li and Zeyi Liu and Pengyu Han and Xiao He and Limin Wang and Tao Zhang},
}

@ARTICLE{MPOS-RVFL,
	author={Han, Pengyu and Chen, Shijin and Liu, Zeyi and He, Xiao},
	journal={IEEE Transactions on Instrumentation and Measurement}, 
	title={Imbalanced Real-Time Fault Diagnosis Based on Minority-Prioritized Online Semi-Supervised Random Vector Functional Link Network}, 
	year={2024},
	volume={73},
	number={},
	pages={1-10},
	doi={10.1109/TIM.2024.3400344}}

@article{wang2025search,
  title={In search of lost online test-time adaptation: A survey},
  author={Wang, Zixin and Luo, Yadan and Zheng, Liang and Chen, Zhuoxiao and Wang, Sen and Huang, Zi},
  journal={International Journal of Computer Vision},
  volume={133},
  number={3},
  pages={1106--1139},
  year={2025},
  publisher={Springer}
}

@article{hu2024cadm+,
  title={CADM+: Confusion-based learning framework with drift detection and adaptation for real-time safety assessment},
  author={Hu, Songqiao and Liu, Zeyi and Li, Minyue and He, Xiao},
  journal={IEEE Transactions on Neural Networks and Learning Systems},
  volume={36},
  number={3},
  pages={5126--5139},
  year={2024},
  publisher={IEEE}
}

@ARTICLE{11104131,
  author={Liu, Zeyi and He, Xiao and Huang, Biao and Zhou, Donghua},
  journal={IEEE Transactions on Cybernetics}, 
  title={Incremental Learning-Enabled Fault Diagnosis of Dynamic Systems: A Comprehensive Review}, 
  year={2025},
  volume={55},
  number={12},
  pages={5633-5649},
  doi={10.1109/TCYB.2025.3586643}}

@ARTICLE{10688394,
  author={Wu, Kangkai and Li, Jingjing and Meng, Lichao and Li, Fengling and Lu, Ke},
  journal={IEEE Transactions on Industrial Informatics}, 
  title={Online Adaptive Fault Diagnosis With Test-Time Domain Adaptation}, 
  year={2025},
  volume={21},
  number={1},
  pages={107-117},
  doi={10.1109/TII.2024.3438240}}

@article{DANN,
author  = {Yaroslav Ganin and Evgeniya Ustinova and Hana Ajakan and Pascal Germain and Hugo Larochelle and Fran{\c{c}}ois Laviolette and Mario March and Victor Lempitsky},
title   = {Domain-Adversarial Training of Neural Networks},
journal = {Journal of Machine Learning Research},
year    = {2016},
volume  = {17},
number  = {59},
pages   = {1-35},
}

@article{MCC5-motor,
  title={Multi-mode Fault Diagnosis Datasets of Three-phase Asynchronous Motor Under Variable Working Conditions},
  author={Chen, Shijin and Liu, Zeyi and Li, Chenyang and Zou, Dongliang and He, Xiao and Zhou, Donghua},
  journal={arXiv preprint arXiv:2601.02278},
  year={2026}
}

\end{document}